\documentclass[twocolumn,amssymb, showpacs, amsmath,superscriptaddress,prl]{revtex4-1}

\usepackage{graphicx}
\usepackage{dcolumn}
\usepackage{bm}

\begin{document}

\title{Testing V$_3$Si for two-band superconductivity}

\author {M. Zehetmayer}
\author {J. Hecher}
\affiliation {Atominstitut, Vienna University of Technology, 1020
Vienna, Austria}

\begin{abstract}
Is V$_3$Si a two-band or a single-band superconductor? Everyone who searches the literature for this question will find conflicting answers, for V$_3$Si was claimed to be a perfect example of two-band and claimed to be a perfect example of single-band superconductivity. In this article we intend to clarify the situation by presenting new experimental facts acquired from the magnetic properties of a V$_3$Si single crystal. We probe for field dependent two-band effects by analyzing the reversible magnetization at different temperatures, and we probe for temperature dependent two-band effects by analyzing the superfluid density obtained by two different methods at different magnetic fields. All our results are reliably described within the single-band models and thus support the single-band scenario for V$_3$Si but do not completely rule out the presence of a very small second gap.

\end{abstract}


\maketitle

\section{Introduction}

Superconductivity in V$_3$Si has been studied since 1953\cite{Har53a} and was considered to be of  conventional s-wave, single-band nature for most of the time. It was only recently that, in the wake of MgB$_2$, V$_3$Si was placed on the list of potential two-band superconductors to explain unconventional experimental results. In particular, its superfluid density was reported  \cite{Nef05a,Kog09a} not only to deviate strongly from the single-band BCS behavior but to match a two-band model, whose interband coupling strength is all but negligible. These conclusions were backed by infrared spectroscopy results and by calculations of the Fermi surface, which was reported to be crossed by several electronic bands \cite{Per10a}. In contrast, a single-band description worked well for the field dependence of the specific-heat and the thermal conductivity \cite{Boa03a}. Furthermore, the field dependence of the magnetic penetration depth and of the vortex core size, determined by muon-spin rotation, were regarded as single-band behavior \cite{Son04b}, and in \cite{Gur04a} the temperature dependence of the specific heat was reported to be conventional. We conclude that we face a confusing situation, which we wished to clarify by further experiments.

V$_3$Si is a member of the A15 superconductors, whose crystal structure changes from cubic at room temperature to tetragonal in the superconducting state \cite{Bat64a}. It is a type II superconductor with a Ginzburg-Landau parameter of about 20, a transition temperature of 16 - 17\,K, and an upper critical field of around 20\,T at 0\,K \cite{Orl79a}; the anisotropy of the superconducting properties is marginal \cite{Khl99a}. The vortex distribution may change from a hexagonal to a cubic lattice as the magnetic field increases \cite{Yet99a}.

In this article we will present new experimental results capable of probing a possible two-band state of V$_3$Si. In the next section, the experimental details and the evaluation methods will be introduced. We will start with the basic characterization of the sample, go on with the measurements of the magnetic moment, and will then show how the reversible magnetization was obtained and fitted using the Ginzburg-Landau model. Finally, the direct determination of the lower critical field will be explained. In the third section, we will present the results. First, we will summarize how two-band effects show up in MgB$_2$ and will then compare this with our findings on V$_3$Si. Potential modifications \cite{Zeh13a} of the field dependencies will be discussed by analyzing the reversible magnetization, those of the temperature dependencies by analyzing the superfluid density and the lower critical field. We will report differences between literature data and our results and present possible reasons. In the final section, we will summarize and once again explain why we believe that V$_3$Si is a single-band superconductor.  

\section{Experiment and evaluation \label{sec:experiment}}

A V$_3$Si single crystal was cut into two pieces of equal sizes of about $3 \times 0.55 \times 0.55$ mm$^3$. Both samples, named VA and VB, became superconducting below 16.7\,K with a transition width of 0.2\,K and had a residual resistance ratio of 33.

Using a SQUID magnetometer, we measured the magnetic moment of sample VB at temperatures from 2 to 16\,K in 1\,K steps as a function of applied magnetic field from 0 to 7\,T. Sample VA was analyzed with a non-commercial rotating sample magnetometer, where the sample is glued on the rim of a circular plate, which rotates at a frequency of 15\,Hz. During one rotation the sample passes four pick-up coils, where it induces electrical voltages proportional to its magnetic moment. For details about the instrument and how the magnetic moment is determined, the reader is referred to \cite{Eis11a}. The main advantage of the rotating sample magnetometer is its fitting into a cryostat with a 15\,T magnet. Accordingly, magnetization loops up to 15\,T were recorded at temperatures of 5.2, 7, 9, 11, 13, and 15\,K.

The resulting magnetization loops, either measured in the SQUID or in the rotating sample magnetometer, revealed reversible behavior over most of the field range. Irreversible effects, caused by flux-pinning, emerged merely at low fields and became dominant near 0\,T. For instance, the critical current density at 5.2\,K, evaluated using the methods presented in \cite{Zeh09a}, decreased with field from some 10$^9$\,Am$^{-2}$ at 0.1\,T to $5 \times 10^7$\,Am$^{-2}$ at 1\,T and to a negligible value at and above 2\,T. Accordingly, more than 85 per cent of the loop fell into the reversible regime. The irreversible effects became slightly larger at lower temperatures but considerably smaller at higher temperatures. The critical currents, which are proportional to the hysteresis width of the magnetization loops, were found to be somewhat larger in the SQUID than in the rotating sample magnetometer, yet the reversible parts agreed well.        

As already mentioned, irreversibility appeared merely at low fields, but even there, knowing the magnetization as a function of increasing, $M(H_{\rm a}^+$), and decreasing applied field, $M(H_{\rm a}^-$), allowed us to determine the reversible part via $M_{\rm r}(H_{\rm a})$ $\simeq$ 0.5[$M(H_{\rm a}^+)$ + $M(H_{\rm a}^-)$], where $H_{\rm a}^+$ and $H_{\rm a}^-$ refer to the same applied field $H_{\rm a}$. This procedure gives reliable results as long as the hysteresis width is not much larger than the corresponding reversible signal. To get $M_{\rm r}(B)$ from $M_{\rm r}(H_{\rm a})$, we calculated the magnetic induction via $B = \mu_0(H - D M_{\rm r} + M_{\rm r})$, with $\mu_0$ = $4 \pi \times 10^{-7}$\,NA$^{-2}$, $H$ the applied field corrected by the field induced by the macroscopic currents \cite{Zeh09a}, and $D$ the numerically calculated demagnetization factor of the sample in the Meissner state. 

The next step was to compare the reversible magnetization, acquired from experiment, with theory. The theoretical magnetization curves of a single-band superconductor were taken from approximate equations of the Ginzburg-Landau theory, provided by Brandt \cite{Bra03a}. According to this paper \cite{Bra03a} the approximation errors should be less than two per cent, for the Ginzburg-Landau parameter of V$_3$Si is large. The Ginzburg-Landau curves depend on just two parameters, namely the upper critical field, at which the magnetization becomes zero, and the Ginzburg-Landau parameter, which determines the shape of the curve. The well-known Ginzburg-Landau relations \cite{Tin75a} allowed us to calculate further quantities, such as the lower and the thermodynamic critical field, the coherence length, the magnetic penetration depth, and the superfluid density.  

The lower critical field was additionally determined by measuring the field at which vortices start to penetrate the sample. After having minimized the stray fields in the SQUID cryostat, we cooled the sample below its transition temperature in zero field. Then we enhanced the applied field stepwise but interrupted each step by measuring the corresponding remanent magnetic moment in zero field. The remanent moment should vanish at low fields and start to increase above the lower critical field, where vortices are created and pinned in the sample.

\section{Results and discussion}

In this section we will analyze whether superconductivity in V$_3$Si is more reliably described by a single or a two-band model. If V$_3$Si is a two-band superconductor, we expect some properties to deviate from the single-band BCS behavior \cite{Zeh13a}. Possible effects on the field dependence will be discussed via the reversible magnetization and possible effects on the temperature dependence via the superfluid density. The results will be compared with the behavior of MgB$_2$, a well-known two-band material. We will see that the single-band models provide a fair description of our results in V$_3$Si.

\begin{figure}
    \centering
    \includegraphics[clip, width = 8.5cm]{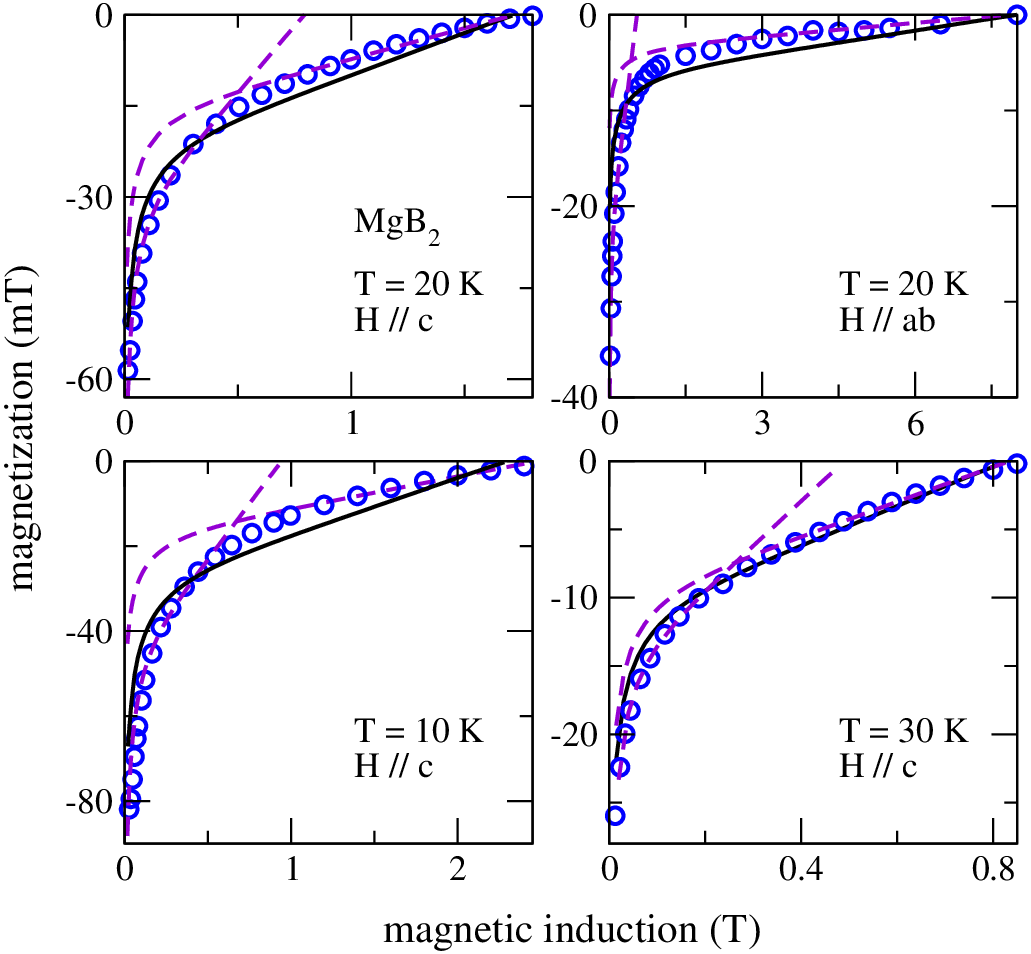}
   \caption{\label{MgB2GLFit} The reversible magnetization of MgB$_2$ as a function of the magnetic induction. The open circles show the experimental data for the applied field oriented parallel ($H \parallel c$) or perpendicular ($H \parallel ab$) to the uniaxial sample axis at 10, 20, and 30\,K (the reduced temperatures are 0.26, 0.51, and 0.77). The solid lines are fits of the single-band Ginzburg-Landau model to the experimental data over the whole field range, while the broken lines are such fits either to the low or to the high field part of the experimental data, reflecting  the two-band nature of MgB$_2$. For details see \cite{Zeh04b}.}
\end{figure}

We start by summarizing some two-band effects of MgB$_2$ \cite{Eis07a}. This material consists of the near-isotropic $\pi$-band and the anisotropic $\sigma$-band, having similar electronic densities of states. Due to interband coupling, the gaps are expected to close at the same field. Yet the superconducting properties of the $\pi$-band are heavily suppressed above a particular field, which is commonly called the upper critical field of the $\pi$-band and whose value is about a third of that of the $\sigma$-band. Accordingly, the field dependence of several superconducting properties deviates significantly from the single-band behavior. This is illustrated in figure~\ref{MgB2GLFit}, where the reversible magnetization of an MgB$_2$ single crystal, indicated by open circles, is shown as a function of the magnetic induction. The solid lines present the single-band Ginzburg-Landau fits. Barring the 30\,K results, the agreement between theory and experiment is poor and the differences are not merely of quantitative but also of qualitative nature. In contrast, the curves can be nicely fitted by {\it two} single-band Ginzburg-Landau curves, as shown by the broken lines in the diagrams. One is adjusted to the low and the other to the high field region of the experiment, thus reflecting the two bands with their different upper critical fields \cite{Zeh04b}. Applying the methods to different field orientations reveals the different anisotropies of the two bands. Similar effects were observed in NbSe$_2$ \cite{Zeh10a}.

\begin{figure}
    \centering
    \includegraphics[clip, width = 8.5cm]{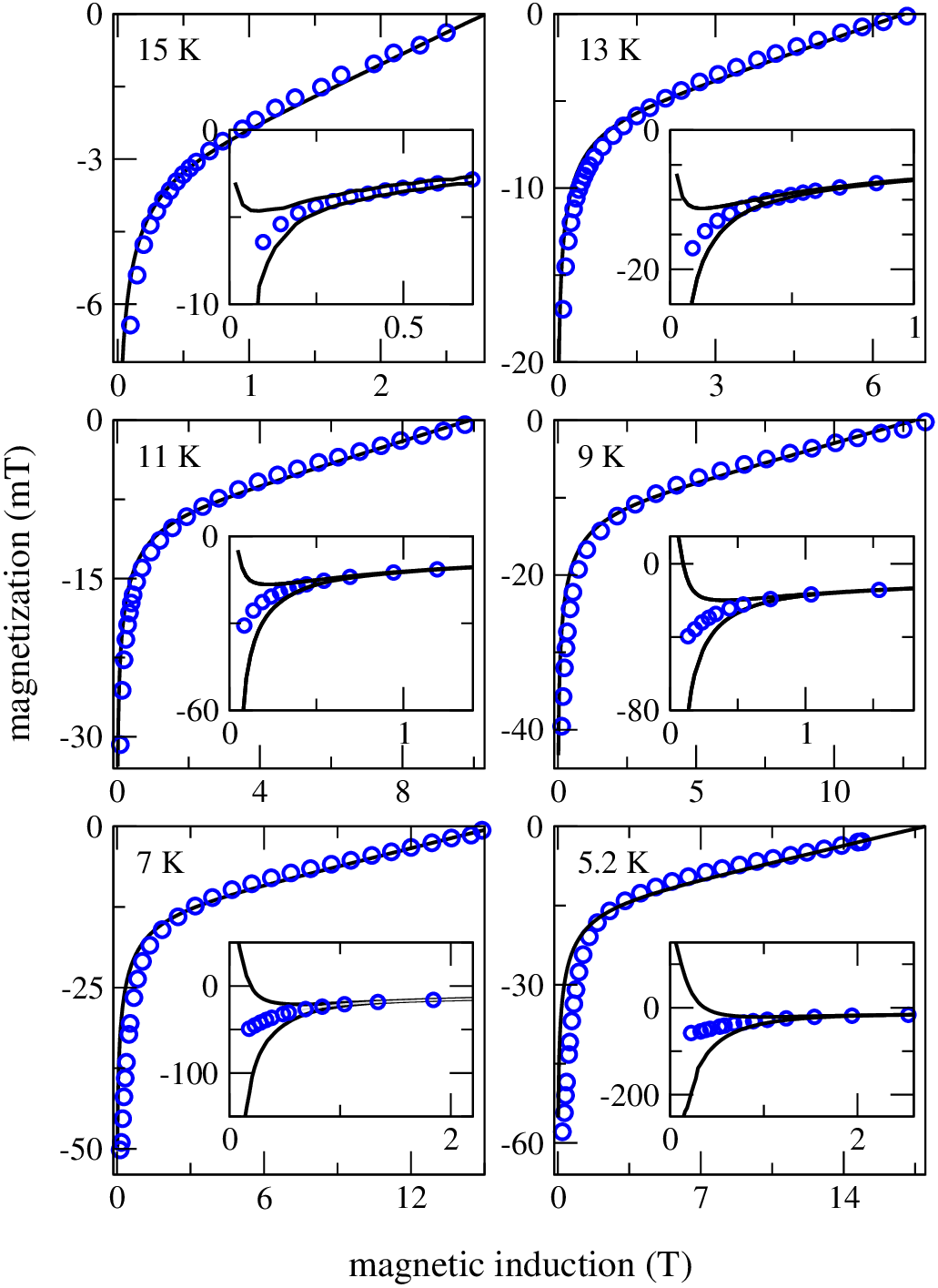}
   \caption{\label{GLFit} The magnetization of V$_3$Si as a function of magnetic induction. The open circles  show the reversible data at 15, 13, 11, 9, 7, and 5.2\,K (the reduced temperatures are 0.90, 0.78, 0.66, 0.54, 0.42, and 0.31) and the solid lines the corresponding Ginzburg-Landau model fits. The solid lines of the insets show the irreversible data.}
\end{figure}

We are now prepared to shift our focus to V$_3$Si. Figure~\ref{GLFit} shows the magnetization as a function of the magnetic induction at temperatures of about 15, 13, 11, 9, 7, and 5\,K. The open circles indicate the reversible magnetization acquired from the rotating sample magnetometer measurements of a V$_3$Si single crystal and the solid lines the single-band Ginzburg-Landau behavior adjusted to the experimental data. In the insets, the solid lines indicate the irreversible magnetization, though only  at low fields, where a significant hysteresis shows up. At not too low temperatures, say about 9 - 15\,K, we consider the agreement between experiment and single-band theory very good. As the temperature is reduced, the differences between theory and experiment get larger,  becoming apparent at low fields, for we adjusted the fits mainly to the high field regions, where the experiments reveal the reversible data directly. 

Next, we will analyze the quality of the fits in more detail. First, we determined the areas under the reversible magnetization curves, which are proportional to the condensation energies. The ratio of the condensation energy obtained from the Ginzburg-Landau fit to that from the experimental curves is considered a sensible measure for the fit quality. In MgB$_2$, the differences in the condensation energies of the high-field fit and the experimental data were some 18 per cent at 30\,K and 30 per cent at 10 and 20\,K. In V$_3$Si, we found much smaller deviations, namely 3 - 8 per cent at 9 - 15\,K and 10 - 12 per cent at 7 and 5.2\,K. So, even the low-temperature fits of V$_3$Si agree with the experimental data better than any single-band fit of MgB$_2$.

There are several possible reasons for the larger deviations between theory and experiment in V$_3$Si at low temperatures. To begin with, as shown in figure~\ref{GLFit}, the low-field irreversible magnetization becomes large at low temperatures, thus enhancing possible errors in the corresponding reversible data. The insets of figure~\ref{GLFit}, however, illustrate that the irreversible magnetization does not become much larger than the reversible data even at low temperatures and small fields, and hence the errors in calculating the reversible magnetization are not serious. On the other hand, we are faced with the shortcomings of the Ginzburg-Landau theory. As this theory is derived from BCS theory in the vicinity of the transition temperature, the potential errors grow when we go to lower temperatures. Yet the theory has been successfully applied to evaluating and describing experimental data at high and low temperatures, as a function of temperature and as a function of field in a large number of publications. In particular, using adjustable parameters instead of the microscopic BCS ones apparently extends the applicability to much lower temperatures (e.g. \cite{Sil12a}). Watanabe et al.~\cite{Wat05a} calculated the field dependence of the reversible magnetization of an s-wave system using the Eilenberger equations, which hold at arbitrary temperatures. We verified that the Ginzburg-Landau model with a Ginzburg-Landau parameter of 49 reproduces the Eilenberger curve of \cite{Wat05a} close to the transition temperature (see also \cite{Bel12a}). Reducing the temperature makes the Eilenberger curves slightly steeper at low fields and slightly flatter at high fields. This is basically what we find for the experimental curves in figure~\ref{GLFit}, namely reducing the temperature makes the experimental curves slightly steeper at low fields and slightly flatter at high fields compared with the Ginzburg-Landau fit. Accordingly, the qualitative deviations between our experiments and the fits are just as expected when we assume that the Eilenberger model describes experiment at all temperatures. The deviations from the Ginzburg-Landau model, assessed via the condensation energies, agree even quantitatively, i.e., we found some 12 per cent for the experimental data and 14 per cent for the Eilenberger curves at a reduced temperature of about 0.3. Granted, the Eilenberger calculations and our experimental data refer to systems with different Ginzburg-Landau parameters ($\kappa$), but both systems belong to the high-$\kappa$ regime and hence should behave similarly. To conclude, the changes of the Eilenberger curves are not substantial when the temperature is lowered, which affirms the usefulness of our approach. Still, we do not expect an exact description at the lowest temperatures of figure~\ref{GLFit}.

In contrast to V$_3$Si, presented in figure~\ref{GLFit}, the experimental data of MgB$_2$, presented in figure~\ref{MgB2GLFit}, also qualitatively disagree with the single-band models. In particular, we found the high-field fit to lead to a much smaller lower critical field, i.e. the magnetization value at $B = 0$\,T, than obtained from experiment via extrapolation of the low-field data, while in V$_3$Si the two curves result in almost the same lower critical fields.

Summarizing, we consider the agreement between single-band theory and experiment in V$_3$Si as good as can be expected in view of the uncertainties within both the theoretical and the experimental data. Accordingly, we conclude that the reversible magnetization of V$_3$Si supports the single-band scenario.

\begin{figure}
    \centering
    \includegraphics[clip, width = 8.5cm]{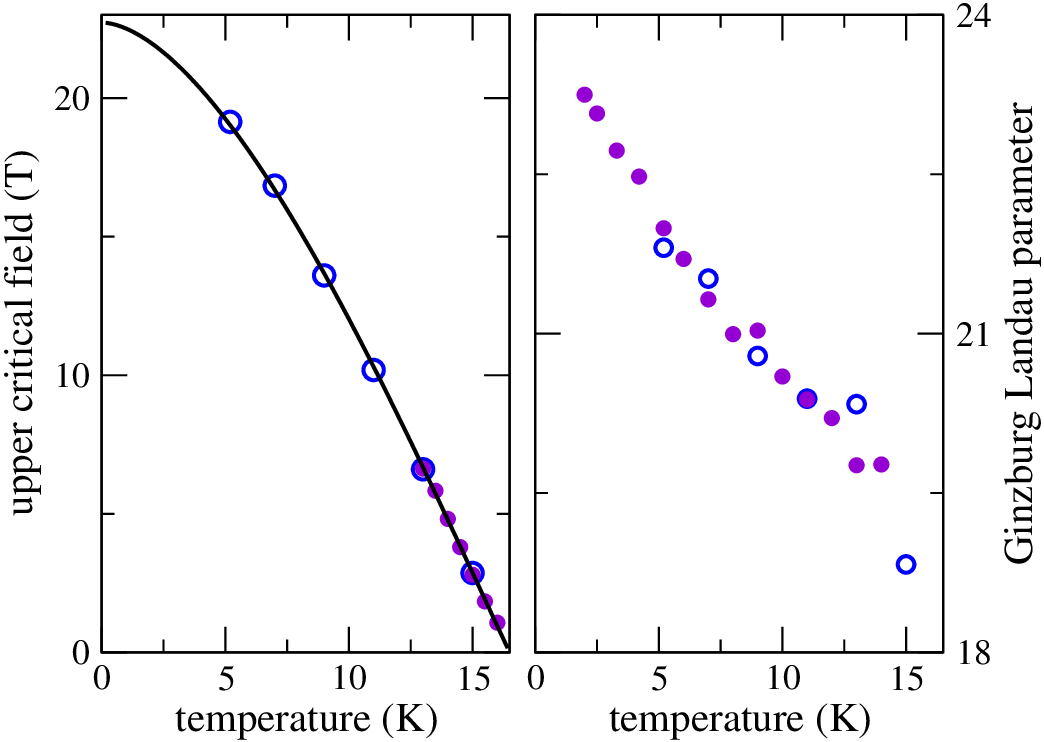}
   \caption{\label{Bc2} The upper critical field, presented in the left panel, and the Ginzburg-Landau parameter, presented in the right panel, of V$_3$Si as a function of temperature. The open circles show experimental data acquired from the rotating sample magnetometer and the full circles experimental data from the SQUID measurements. The upper critical field follows the clean-limit BCS behavior, depicted by the solid line, leading to 22.7\,T at 0\,K. The Ginzburg-Landau parameter decreases roughly linearly from about 24 to 19 as the temperature increases from 0\,K to the transition temperature.}
\end{figure}

Figure~\ref{Bc2} presents the upper critical field and the Ginzburg-Landau parameter of V$_3$Si. The open circles indicate results from the rotating sample and the full circles those from the SQUID magnetometry. The upper critical field has been evaluated by adjusting the Ginzburg-Landau model merely to the high-field regime of the reversible magnetization and may thus slightly deviate from the results obtained from full range fits. The solid line presents the clean-limit single-band BCS behavior \cite{Hel66a}, is in excellent agreement with our experimental data, and leads to 22.7\,T at 0\,K, which matches literature data well \cite{Orl79a}. In contrast to many two-band materials, no clear upward curvature near the transition temperature appears \cite{Shu98a,Zeh02a,Mun12a}. The Ginzburg-Landau parameter, taken from fits over the whole field range and presented in the right panel, decreases quite linearly from about 24 at 0\,K to 19 at the transition temperature, a behavior that is also close to the single-band BCS prediction \cite{Hel66a}. Calculating further properties employing the Ginzburg-Landau relations \cite{Tin75a} resulted in about 90\,nm for the magnetic penetration depth, 4\,nm for the coherence length, 0.6\,T for the thermodynamic critical field, and 0.07\,T for the lower critical field at 0\,K. The critical lengths are close to literature data \cite{Son04b,Yet05a}.

\begin{figure}
    \centering
    \includegraphics[clip, width = 8.5cm]{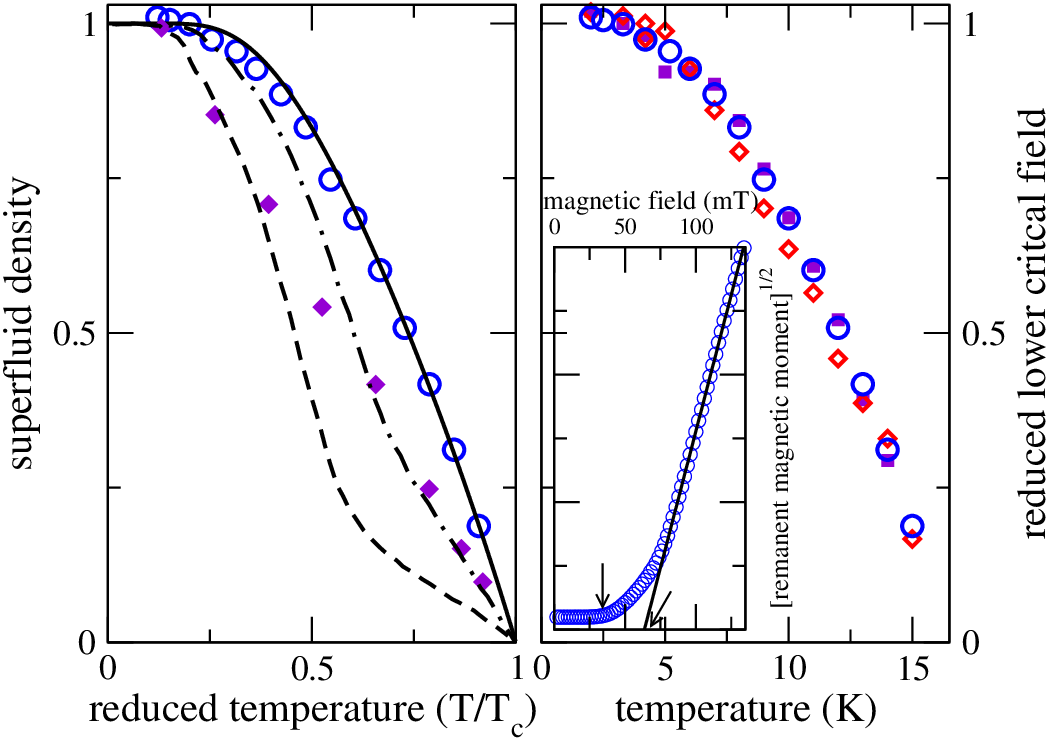}
   \caption{\label{SFLD} The superfluid density as a function of reduced temperature (left panel) and the reduced lower critical field as a function of temperature (right panel). In the left panel, the open circles show our experimental result on V$_3$Si, obtained from the reversible magnetization, which is in good agreement with the single-band BCS behavior, indicated by the solid line. The broken and the dot-dashed lines show the two-band-like superfluid density of V$_3$Si reported in Refs.~\cite{Nef05a} and \cite{Kog09a} schematically; the full diamonds illustrate the two-band behavior of MgB$_2$ \cite{Zeh04b}. In the right panel, the open circles show the reduced lower critical field of our V$_3$Si single crystal, evaluated from the reversible magnetization, which is all but identical to the superfluid density of the left panel. The open diamonds and the full squares are obtained from direct measurements of the first vortex-penetration field as indicated by the arrows in the inset. The symbols in the inset present the square root of the remanent magnetic moment as a function of applied field at 7\,K; the solid line is a linear fit to the high-field data.}
\end{figure}

We proceed by analyzing the temperature dependent effects by means of the superfluid density. Having evaluated the upper critical field, $B_{\rm c2}$, and the Ginzburg-Landau parameter, $\kappa$, via the above fit procedure, we acquired the magnetic penetration depth, $\lambda$, by using the Ginzburg-Landau relations, and the superfluid density, $\rho_{\rm s}$, via $\rho_{\rm s}$(T) = [$\lambda$(0\,K) / $\lambda$(T)]$^2$, with $\lambda = \kappa \xi$, $\xi^2 = \Phi_0 / (2 \pi B_{\rm c2})$, and $\Phi_0 \simeq 2.07$ $\times 10^{-15}$\,Vs. To assess the penetration depth at 0\,K, we used the SQUID measurements, where the temperature could be reduced to 2\,K, but the magnetic fields were limited to 7\,T. We therefore took the inter- and extrapolated upper critical fields from the rotating sample magnetometer results for fitting the SQUID data, so that only the Ginzburg-Landau parameter remained to be determined. To justify the use of these data, we verified that the reversible curves from SQUID and rotating sample magnetometer agree in the overlapping field and temperature range.

The left panel of figure~\ref{SFLD} presents the superfluid density of V$_3$Si, indicated by open circles, as a function of reduced temperature. The solid line shows the expected behavior of a single-band BCS superconductor, which is close to our experimental data. Figure~\ref{SFLD} shows also the superfluid density of MgB$_2$, indicated by full symbols, as an example for two-band superconductivity. In comparison with V$_3$Si, the MgB$_2$ curve decreases much faster at low temperatures and then becomes almost linear as the temperature increases. To explain this behavior, we need to consider that a two-band superconductor has two distinct energy gaps. The smaller gap reduces the excitation energy on the corresponding part of the Fermi surface and hence makes the superfluid density decrease more rapidly at small temperatures. This may lead to a near-linear behavior at intermediate temperatures, as found in several two-band materials, such as MgB$_2$, NbSe$_2$, and the iron-based superconductors \cite{Man02a,Fle07a,Mar09b}. We conclude that also the superfluid density of V$_3$Si supports the single-band scenario.

We now come back to the two-band scenario of V$_3$Si proposed in Refs.~\cite{Nef05a} and \cite{Kog09a}, based on measurements of the superfluid density. The broken line in figure~\ref{SFLD} shows the result presented in \cite{Nef05a}, which was obtained by measuring the microwave surface impedance of a single crystal. There is no doubt that this curve is totally different from our result, that is, it decreases more rapidly at low temperatures and has a sharp kink at about 0.6\,$T_{\rm c}$. Analyzing their data with a two-band model, the authors found distinctly different energy gaps, though similar intraband coupling strengths for the two bands and, as indicated by the sharp kink at intermediate temperatures, almost negligible interband coupling. The superfluid density published by Kogan et al.~\cite{Kog09a}, shown schematically in figure~\ref{SFLD} by the dot-dashed line, was measured by a tunnel diode resonator technique and is quite different from that of \cite{Nef05a}, yet it also reflects a two-band scenario with similar intraband and near-negligible interband coupling strengths. 

What are the possible reasons for the differences between those literature \cite{Nef05a,Kog09a} and our data? Two points are obvious. First, while we evaluated the penetration depth from fits to the reversible magnetization over the whole field range, save for very small fields, where the experimental data are not available or not reliable, the authors of Refs.~\cite{Nef05a} and \cite{Kog09a} evaluated their data solely at very low fields. Thus, a second band, but one with a very small upper critical field, would resolve the contradictions. The second point is that while our method probes the bulk, the methods of Refs.~\cite{Nef05a} and \cite{Kog09a} probe the sample surface. Thus, surface irregularities or inhomogeneities that may change the properties on the surface would also resolve the contradictions. Such a statement, however, remains a speculation, for we know nothing about the surfaces of the samples used in the studies. Diener et al.~\cite{Die09a} faced that problem in MgCNi$_3$. Acquiring data with the same method as used in \cite{Kog09a} resulted in a penetration depth behavior similar to the broken curves of figure~\ref{SFLD}, while acquiring the results from measurements of the lower critical field resulted in a BCS-like behavior. The authors suggested that those differences might be caused by inhomogeneities at the sample surface and hence considered the BCS-like behavior correct. 

We also determined the superfluid density at very low magnetic fields,  namely by measuring the lower critical field directly. But measuring the lower critical field directly is anything but trivial. To begin with, we can merely determine the field at which flux lines start to penetrate into the sample. This is accomplished by recording the remanent magnetic moment of the sample in zero field as a function of the maximum applied field, as described at the end of Sec.~\ref{sec:experiment}. Unfortunately, this first penetration field is usually not simply connected with the lower critical field via a single-valued demagnetization factor. First, sample edges give rise to very high stray fields, which may surpass the lower critical field and hence enforce the creation of vortices at very low applied fields, but we do not know well how the local induction is related to the applied field in such a configuration. Second, surface irregularities would affect the creation of vortices, while clean surfaces may induce additional barriers. 

So, how can we acquire useful results from that procedure? To begin with, we are mainly interested in the relative temperature dependence of the lower critical field and not so much in its absolute value. Second, we aligned the longest length of our sample, the size of which is $3 \times 0.55 \times 0.55$ mm$^3$, parallel to the applied field, so that the geometry effects were as small as possible and assumed those geometry effects to be temperature independent. As expected, when we increased the applied field, the remanent magnetic moment remained zero at the beginning, then started to rise at a slope that became gradually steeper, indicating that first vortices had been formed and pinned in the sample, and eventually increased quadratically with field (see inset of figure~\ref{SFLD}). Assuming a constant current density parallel to the sample borders and ignoring the stray fields revealed this quadratic behavior also in calculations (cf. with \cite{Mos91a}). We thus consider the extrapolated onset of the quadratic behavior (cf. with inset of figure~\ref{SFLD}) as a reliable assessment of the lowest field where vortices parallel to the applied field were in the sample. 

We determined both the smallest field at which we observed a slope in the remanent magnetic moment and the field obtained from extrapolating the quadratic part to zero (see inset of figure~\ref{SFLD}). The temperature dependence of both was found in good agreement with the superfluid density acquired from the Ginzburg-Landau fits, as shown in the right panel of figure~\ref{SFLD}. Note that Ginzburg-Landau theory predicts the same temperature dependence for the superfluid density and the lower critical field when the changes in the logarithm of the Ginzburg-Landau parameter can be ignored, as is the case in our sample. To assess the absolute lower critical field values, we multiplied the penetration fields with the factor $(1-D)^{-1}$ $\simeq$ 1.15, where $D \simeq 0.13$ is the averaged demagnetization factor. In comparison to the lower critical fields from the Ginzburg-Landau fits, leading to 68\,mT at 0\,K, this led to lower values by some 10 per cent, namely 60\,mT at 0\,K, from evaluating the onset fields, and to larger values by 40 per cent, namely 95\,mT at 0\,K, from evaluating the fields where the quadratic behavior started. We consider these differences to be reliable. Finally, we conclude that also at very low magnetic fields, the temperature dependence shows no indication of a two-band behavior.

Having presented clear support for the single-band behavior, we may ask if our results rule out the two-band scenario completely. This is not the case, for our experiments would not detect a second band if its contribution to the measured quantities were marginal or if its superconducting properties were all but identical to the first band. These scenarios, however, would be different from those proposed in literature \cite{Nef05a,Kog09a}. Which properties would a hypothetical second band have in our sample? Above, we have discussed the fit quality of the reversible magnetization data in terms of the ratio of the condensation energies from the fit and from experiment. In MgB$_2$ the differences, which are some 30 per cent at low temperatures, are ascribed to the two-band scenario. Accordingly, 70 per cent of the condensation energy are induced by the $\sigma$-band, which is the band probed by the high-field fits \cite{Zeh04b}, and 30 per cent by the second band, the $\pi$-band, {\it and} interband coupling effects, in rough agreement with \cite{Eis05a}. In V$_3$Si, the deviations are much smaller and we have shown that they can qualitatively and quantitatively be explained by the imperfections of the Ginzburg-Landau model. Nevertheless, we analyzed the results assuming the deviations to be caused by a hypothetical second band. For example, at 9\,K the fit error was 5 per cent and hence the second band would contribute to the condensation energy ($E_{\rm c}$) by {\it less} than 5 per cent, for these 5 per cent include the interband contributions. Employing BCS theory, where $E_{\rm c}$ = $0.5 N \Delta^2$, and assuming both bands to have the same density of states ($N$) would result in a gap ($\Delta$) ratio larger than 4.4 : 1 (1 : $\sqrt{5/95}$), which would be quite large and much larger than the ratios proposed in Refs.~\cite{Nef05a,Kog09a}, namely 1.4 - 1.8 : 1. 

Strong interband effects would also mask two-band effects \cite{Nic05b}, though it should be noticed that the two-band scenario for V$_3$Si \cite{Nef05a, Kog09a} came along with a negligible interband coupling strength and the high purity of our single crystals makes strong interband impurity-scattering unlikely. 

\section{Summary and conclusion}

Let us sum up what we have learned from testing V$_3$Si for two-band superconductivity. We have shown that the field dependence of the reversible magnetization matches the single-band Ginzburg-Landau theory reliably well. The differences between the experimental data and the theoretical fits grow as the temperature is reduced but remain small and can qualitatively and quantitatively be explained by the imperfections of the Ginzburg-Landau model. We have also found that the temperature dependence of the superfluid density, determined with different methods at different magnetic fields, follows single-band BCS behavior. Accordingly, all our results support a single-band scenario for V$_3$Si. 

\section*{Acknowledgments}
This work was supported by the Austrian Science Fund under Contract No. 21194
and 23996.

\end{document}